%% file: main.tex
\documentclass{article}

\newif\ifarXiv
\arXivtrue

\ifarXiv
  \usepackage{natbib}
  \usepackage{authblk}

  \usepackage[dvipsnames]{xcolor}  

  \author[1,2,3]{Leon Hetzel$^*$}
  \author[1,3]{Johanna Sommer$^*$}
  \author[1,2]{Bastian Rieck}
  \author[1,2,3]{Fabian Theis}
  \author[1,3]{Stephan~Günnemann\vspace{-0.5em}}

  \affil[$\phantom{M}$]{\texttt{\small{\{l.hetzel, jm.sommer, s.guennemann\}@tum.de}}}
  \affil[$\phantom{B}$]{\texttt{\small{\{bastian.rieck, fabian.theis\}@helmholtz-munich.de}\vspace{0.5em}}}
  \affil[1]{School of Computation, Information, and Technology, Technical University of Munich}
  \affil[2]{Helmholtz Munich}
  \affil[3]{Munich Data Science Institute, Technical University of Munich}

  \usepackage[letterpaper]{geometry}

  \date{}

  \usepackage{mathpazo}
\else
    \PassOptionsToPackage{sort}{natbib}
    \PassOptionsToPackage{dvipsnames}{xcolor}
    \usepackage{iclr2024_conference}
    \usepackage{times}
\fi

\newcommand{\parencite}[1]{\citep{#1}}
\newcommand{\textcite}[1]{\citet{#1}}

\usepackage[utf8]{inputenc} 
\usepackage[T1]{fontenc}    
\usepackage{hyperref}       
\usepackage{url}            
\usepackage{booktabs}       
\usepackage{amsfonts}       
\usepackage{nicefrac}       
\usepackage{microtype}      

\usepackage{multirow}
\usepackage{adjustbox}
\usepackage{array}
\usepackage{booktabs}
\usepackage{multirow}
\usepackage{wrapfig}
\usepackage{paralist}
\usepackage{subcaption}
\usepackage{floatrow}
\usepackage[british]{babel}

\usepackage{array,graphicx}
\usepackage{booktabs}
\usepackage{pifont}

\usepackage{amsmath}
\usepackage{dsfont}
\usepackage{subcaption}
\usepackage{graphicx}
\usepackage[capitalise]{cleveref}

\hypersetup{
  breaklinks,
  colorlinks,
  linkcolor=OrangeRed,
  citecolor=RoyalBlue,
  urlcolor=RoyalBlue,
}

\newcommand{\modelname}{MAGNet}

\title{\modelname: Motif-Agnostic Generation \\ of Molecules from Shapes}

\ifarXiv
\else
  \author{%
  }
\fi

\begin{document}

\newcommand{\figref}[1]{Fig.~\!\ref{#1}}
\newcommand{\sG}{\ensuremath{\mathcal{G}}}
\newcommand{\sL}{\ensuremath{\mathcal{L}}}
\newcommand{\sC}{\ensuremath{\mathcal{C}}}
\newcommand{\sA}{\ensuremath{\mathcal{A}}}
\newcommand{\sS}{\ensuremath{\mathcal{S}}}
\newcommand{\sM}{\ensuremath{\mathcal{M}}}
\newcommand{\sJ}{\ensuremath{\mathcal{J}}}
\newcommand{\prob}{\ensuremath{\mathbb{P}}}
\newcommand{\qprob}{\ensuremath{\mathbb{Q}}}
\newcommand{\card}[1]{\lvert #1 \rvert}

\maketitle

\begin{abstract}
  \input{chapters/0_abstract}
\end{abstract}
\ifarXiv
    \footnotetext{$^\ast$ These authors contributed equally}
    \footnotetext{$\phantom{^\ast}$ Code is available at \href{https://github.com/TUM-DAML/MAGNet}{github.com/TUM-DAML/MAGNet}}
\else
\fi
\input{chapters/1_introduction}
\input{chapters/2_modelling}
\input{chapters/3_relatedwork}
\input{chapters/4_experiments}
\input{chapters/5_conclusion}


\bibliography{ref}
\bibliographystyle{iclr2024_conference}

\appendix
\input{chapters/6_appendix}

\end{document}

%% file: chapters/0_abstract.tex
Recent advances in machine learning for molecules exhibit great potential for facilitating drug discovery from \emph{in silico} predictions.
Most models for molecule generation rely on the decomposition of molecules into frequently occurring substructures (motifs), from which they generate novel compounds. 
While motif representations greatly aid in learning molecular distributions, such methods struggle to represent substructures beyond their known motif set. 
To alleviate this issue and increase flexibility across datasets, we propose \modelname, a graph-based model that generates abstract shapes before allocating atom and bond types. 
To this end, we introduce a novel factorisation of the molecules' data distribution that accounts for the molecules' global context and facilitates learning adequate assignments of atoms and bonds onto shapes. Despite the added complexity of shape abstractions, \modelname\ outperforms most other graph-based approaches on standard benchmarks. Importantly, we demonstrate that \modelname's improved expressivity leads to molecules with more topologically distinct structures and, at the same time, diverse atom and bond assignments.




%% file: chapters/1_introduction.tex
\section{Introduction}
The role of machine learning (ML) models in generating novel compounds has grown significantly, finding applications in fields like drug discovery, material science, and chemistry \parencite{bian_generative_2021, butler_machine_2018, choudhary_recent_2022, hetzel_predicting_2022, moret_leveraging_2023}. These models offer a promising avenue for efficiently navigating the vast chemical space and generating unique molecules with specific properties \parencite{zhou_optimization_2019, hoffman_optimizing_2022}.
A key ingredient contributing to the success of these models is their ability to encode molecules in a meaningful way, often employing graph neural networks (GNNs) to capture the structural intricacies \parencite{gilmer2017neural, shi_masked_2021, mercado_graph_2021}.
Moreover, the inclusion of molecular fragments, known as motifs, significantly influences the generation process by enabling the model to explicitly encode complex structures such as cycles \parencite{sommer_power_2023}. This contrasts with the gradual formation of ring-like structures from individual atoms, forming chains until both ends unite.

In the context of fragment-based models, researchers have adopted various techniques to construct fragment vocabularies, which can be categorised into chemically inspired and data-driven approaches. For example, both JT-VAE \parencite{jin_junction_2018} and MoLeR \parencite{maziarz_learning_2022} adhere to a heuristic strategy that dissects molecules into predefined structures, primarily consisting of ring systems and acyclic linkers. However, the diverse appearances of molecular structures result in various challenges concerning the vocabulary. While the large fragments in JT-VAE do not generalise well to larger datasets, the number of fragments becomes an issue for MoLeR. Such challenges are not unique to heuristic fragmentation methods but also extend to data-driven approaches like PS-VAE \parencite{kong_molecule_2022b} and MiCaM \parencite{geng2022novo}. These approaches can set the number of chosen fragments but often resort to representing cyclic structures by combinations of chains. MiCaM, in particular, takes an approach that additionally incorporates attachment points for each fragment, resulting in a vocabulary that is ``connection-aware'', increasing its size by a significant margin.
%
%
This leads to a situation where the included fragments often fall short to comprehensively represent the full spectrum of molecules present in the datasets \parencite{sommer_power_2023}. Consequently, a generative model must refer to individual atoms in order to generate uncommon structures, a demanding task as the infrequent structures are often also more complex.

Our work addresses this issue by abstracting fragments to shapes, representing only the binary adjacencies, and using this abstraction for vocabulary creation. Since the same shape can lead to multiple distinct atom and bond representations, this effectively reduces the required vocabulary size to capture a dataset's molecules, while still encoding any hard-to-learn structural elements. Simultaneously, we propose an effective generation procedure in the form of our \modelname\ model, which learns to generate molecules from abstract shape sets. The shape abstraction enables us sample the \emph{entire} molecule context at each stage and construct the molecular graph in a hierarchical fashion. This hierarchy progresses from shapes and their connectivity to the atom level, where shape representations are generated, ultimately leading to the full definition of the molecule. Notably, our model is the first to freely learn the distribution over shape representations, enabling it to sample a greater variety of atom and bond attributes compared to fragment-based approaches.

%
While we identify limitations in some of the benchmarks
for \emph{effectively} evaluating whether complex structures have been captured, we provide alternative analyses to offer a more comprehensive understanding of the generative behaviour of our and other approaches.
Overall, our results highlight \modelname's ability to reliably reconstruct uncommon shapes, including large cycles, surpassing other methods that face challenges in this area. 
Additionally, \modelname\ excels in sampling a wider range of shape representations, closely aligning with the original data distribution.
We also demonstrate \modelname's advantageous properties, including its fidelity to the presented latent code---a departure from existing methods.
Finally, our model enables simultaneous conditioning on multiple scaffolds, underscoring its versatility for diverse applications in molecular generation.
\begin{figure}[t]
\centering
\includegraphics[width=0.9\textwidth]{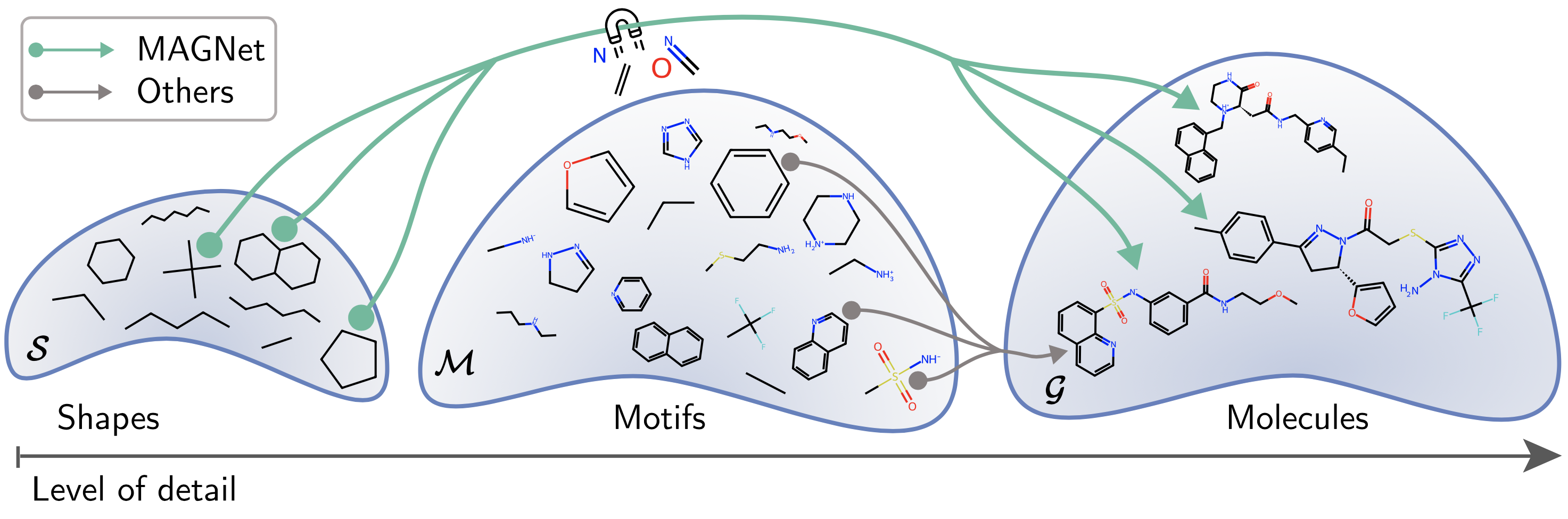}
\caption{Current methods use motifs to compose new molecules. \modelname\ advances this perspective and abstracts molecular subgraphs to untyped shapes. Being agnostic to atom and bond features in the beginning of the generation process, \modelname\ generates molecules with great structural variety.} 
\label{fig:herofig}
\end{figure}

%% file: chapters/2_modelling.tex
\section{Modelling molecules from shapes}
%
We will first detail the factorisation of the data distribution of molecular graphs, denoted as $\prob(\sG)$. Following this, we will introduce our novel approach to fragmenting molecules into shapes and present the corresponding \modelname\ model.

\paragraph*{Factorising the data distribution $\prob(\sG)$}
\label{sec:pG_factorisation}

A molecular graph $\sG$ is defined by its structure together with its node and edge features, describing atoms and bonds, respectively. In this work, we consider a factorisation of $\prob(\sG)$ that builds the molecular graph $\sG$ from its shape representation, i.e.\
\begin{alignat*}{3}
    \prob(\sG) &= \prob(\sG \mid \sG_\sS)\,\prob(\sG_\sS) \: ,
\end{alignat*}
where $\sG$ refers to the normal molecular graph and $\sG_\sS$ to its coarser shape graph.
To factorise $\prob(\sG_\sS)$ further, we consider a fragmentation into a multiset of shapes $\sS$ and their typed connectivity $A \in \{0,\text{C},\text{N},\dots\}^{\card{\sS}\times \card{\sS}}$. Each shape $S\in\sS$ can be classified as one of three categories, namely
\begin{inparaenum}[(i)]
\item rings,
\item junctions, and
\item chains.
\end{inparaenum}
A shape $S$ only holds \emph{structural information} about its $s=\card{S}$ nodes, meaning we can represent it as a binary matrix, i.e.\ $S\in\{0,1\}^{s \times s}$. The shape connectivity $A$, on the other hand, is typed and encodes whether two shapes share an atom, signified by $A\neq0$, and also the respective atom type of the join, e.g.\ C or N. Note that $A$ is a shape-level representations which does not hold any information about the exact position of the connection between shapes. Further, since all cyclic structures within the molecular graph are considered individual shapes, $A$ always describes a tree on the shape-level. 

Moving forward to the atom-level representation, a shape $S$ can be equipped with node and edges features, corresponding to atom and bond types, respectively. We consider this representation of $S$ to constitute a typed subgraph $M$, or fragment, of the input graph $\sG$, and the associated multiset within the molecule $\sM$. We note that a shape $S$ can map to multiple distinct $M$ as only the binary adjacencies will be shared between them; we will later on make use of context knowledge to select a suitable~$M$.

The shape representations $\sM$ and the shape connectivity $A$ do not fully describe $\sG$ yet as they do not specify the exact atomic positions where the shapes have to be joined. For this, we define the join set $\sJ$ which can be understood as the set of join nodes $j$ that are contained in two motifs: $\sJ= \{j\mid j\in M_k,\ j \in M_l,\ A_{kl}\neq 0 ,\ S_k,S_l \in \sS \}$. Finally, the full molecule is defined by attaching leaf atoms $\sL$, which are those atoms with degree $d_i=1$ and with neighbour $j\in \mathcal{N}_i$ with $d_j=3$. 

In conclusion, $\prob (\sG)$ is factorised into the shape representations $\sM$, the join set $\sJ$, and the leaf set $\sL$: 
\begin{alignat*}{3}
    \prob(\sG)  &= \prob(\sL, \sJ, \sM) \\
                &= \prob(\sL \mid \sM, \sJ)\prob(\sJ \mid \sM, A)\,\prob(\sM \mid \sG_\sS)\,\prob(\sG_\sS) \:, 
\end{alignat*}
where we use $\sG_\sS$ to factorise the distribution.  Note that $\sJ$ is \emph{conditionally independent} of $\sS$ given the motifs $\sM$. The same applies to $\sL$ being conditionally independent of $A$ given $\sJ$.

During learning, we start with the coarser shape-level, that is $\sS$ and $A$, and then proceed to the atom level for $\sM$ and $\sJ$. Finally, we learn the distribution of leaves $\sL$. Note that unlike sequential generation methods, this factorisation considers the entire molecule context at all times with only the level of detail increasing.

\subsection{Fragmentation into shapes}
\begin{figure}
\centering
\includegraphics[width=0.9\textwidth]{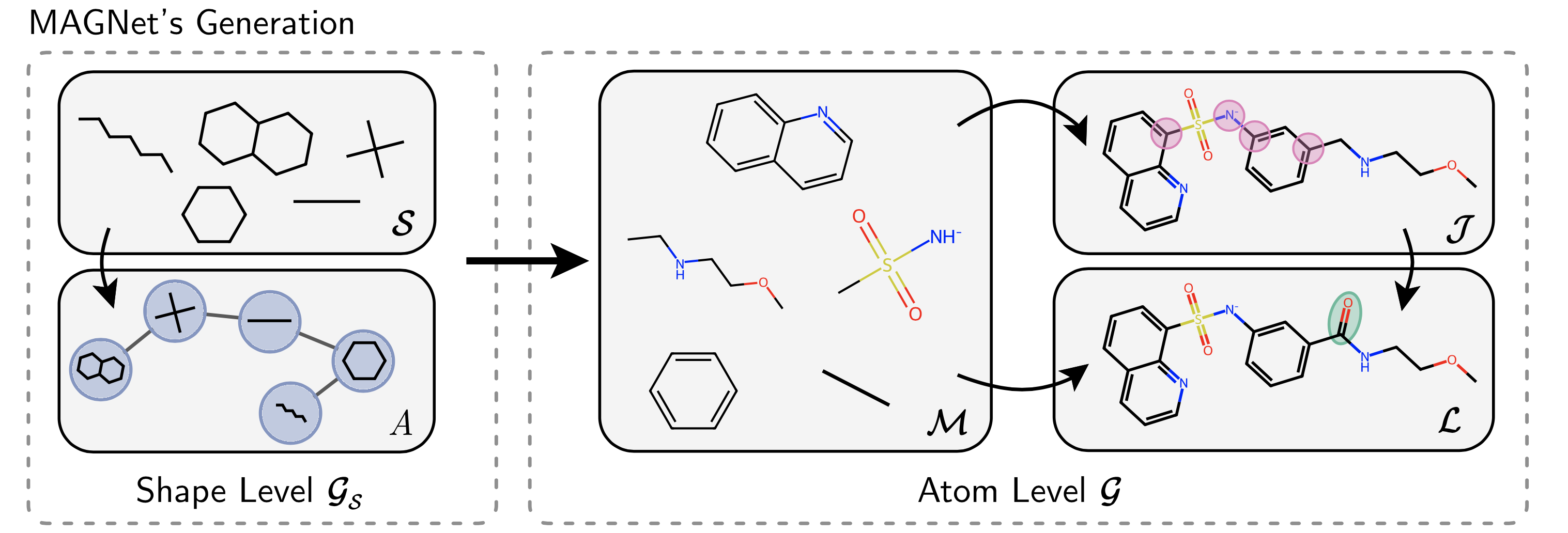}
\caption{On the shape-level, \modelname\ predicts the shape multiset \sS\ and its connectivity $A$. Progressing to the atom-level, $\sG_\sS$ informs the generation of shape representations \sM. To fully define the molecular graph $\sG$, the joining positions \sJ\ and the leaf atoms \sL\ are predicted.} 
\label{fig:decoding_flow}
\end{figure}
\modelname's fragmentation scheme aims to break down a given molecule into clear structural elements. Initially, we remove all leaf atoms $\sL$ across the graph, following the approach outlined in previous works \parencite{jin_hierarchical_2020, maziarz_learning_2022}. Leaves are those nodes that have degree $d_i = 1$ and whose neighbouring node $j\in\mathcal{N}_i$ fulfills $d_{\mathcal{N}_i} = 3$. This step helps to divide the molecule $\sG$ into cyclic and acyclic parts. Importantly, instead of modelling the connection between two fragments $M_i$ and $M_j$ with a connecting bond, we represent it by a shared atom, referred to as the join atom $\nu \in \sJ$. One benefit of this approach is that fragments are not truncated and, for example, two cycles which are connected by only one join atom are identified as two cycles instead of a single cycle that is connected to a chain at both ends.

Moreover, in order to reduce the number of required shapes as much as possible, we further decompose the resulting acyclic fragments. To this end, we introduce ``junctions'' which are defined by a center node with degree three or four present in an acyclic structure. The junction then contains the center nodes as well as its neighbours. This additional refinement reduces the number of fragments by a factor of three, allowing to classify them into rings, junctions, and chains. On the ZINC dataset~\parencite{irwin_zinc20_2020}, consisting of 249,456 compounds, our fragmentation results in 7371 typed subgraphs. When compared to the decomposition BBB procedure \parencite{jin_hierarchical_2020, maziarz_learning_2022}, our approach reduces complexity by collapsing acyclic structures into distinct shapes, leading to a reduction in vocabulary size by more than a half. Additionally, in contrast to data-driven methods like those outlined in \textcite{kong_molecule_2022b} and \textcite{geng2022novo}, our decomposition method maintains structural integrity through a top-down approach.

Abstracting these typed subgraphs further to shapes ultimately results in 347 distinct shapes, where---in the best case---up to almost 800 fragments are consolidated into a single shape token. While the generative model is required to map a single shape to all its various representations, this fragmentation also enables us to model smoother transitions between shape representations. For instance, consider the molecules ``C1NNCC1'' and ``C1NCNC1'', which only differ in the relative positioning among atoms. Using shapes, the model can account for this difference without selecting potentially dissimilar tokens from a large vocabulary. We discuss the potential of our approach for transferability across datasets in \cref{sec:exp_flexibility} and \cref{app:transferability_vocab}.
\subsection{\modelname's generation process}
\label{sec:decoder}
\modelname\ is designed to represent the hierarchy of the factorisation into shape- and atom-level from \cref{sec:pG_factorisation}, cf. \figref{fig:decoding_flow}. The model is trained as a VAE model \parencite{kingma_variational_2015}, where the latent vectors $z$ are trained to encode meaningful semantics about the input data. That is, given a vector $z$ from the latent space, \modelname's generation process first works on the shape level to predict $\sG_\sS$, defined by the multiset \sS\ its connectivity $A$, before going to the atom level which is defined by the fragments \sM, joins \sJ, and leaves \sL.
%
%
\paragraph{Shape-Level}
On the shape-level, \modelname\ first generates the \textbf{shape multiset \sS}---the same shape can occur multiple times in one molecule---from the latent representation $z$. More specifically, we learn $\prob(\sS\mid z)$ by conditioning the generation on the latent code $z $ and the intermediate representation of the shapes by a transformer model. On the sorted shape set, the network is optimised via Cross-Entropy (CE) loss, denoted by $L_\sS$.

Given the shape multiset \sS, \modelname\ infers the \textbf{shape connectivity $A$} between shapes $S_i, S_j \in \sS$. Formally, we learn $\prob(A \mid \sS, z) = \prod_{i,j=1}^{n}\prob(A_{ij}=t\mid S, z)$ where $t\in \{0,\text{C},\text{N},\dots \}$ not only encodes the existence~(or~absence) of a shape connection but also its atom type.
Further, we assume the individual connections, $A_{ij}$ and $A_{lk}$, to be independent given the shape multiset $\sS$ and the latent representation $z$. \modelname\ implements this connectivity module using an MLP, optimised with CE loss $L_A$. We compute the loss on the shape level as $L_{\sG_\sS}=L_\sS + L_A$. It is worth noting that the same shape can have different atom representations depending on its positions in the molecular graph, highlighting the importance of predicting $A$ for generating atom-level shape representations.
%
%
\paragraph{Atom-Level}
Leveraging the shape set \sS\ and connectivity $A$, which together define a molecule's shape-level representation, \modelname\ transitions to the atom-level by discerning appropriate node and edge attributes for each shape, referred to as \textbf{atom and bond types \sM}. This step is pivotal in \modelname's generation process, granting it independence from a pre-established fragment set and thus enhancing its flexibility.
To model the shape representation $\prob(M_i \mid \sS, A, z)$ of shape $S_i$, the atoms are generated using a transformer model that constructs its memory by encoding the shape graph and incorporating embeddings tailored to each individual shape \parencite{shi_masked_2021}. Unlike the shape set, the sizes of the shapes are fixed, enabling a structured representation of each position within the shape. 

Subsequently, the resulting atom embeddings are leveraged to determine the corresponding \textbf{bond types $M^b$} between connected nodes. This bond determination is achieved through the utilisation of an MLP, which effectively captures the necessary relationships for assigning correct bond types.
Note that conditioning on $A$ ensures that $M_k$ includes all atoms required for connectivity also on the atom-level, i.e. the atom allocations for $M$ have to respect all join types defined by the shape connectivity $A$: $A_{kl} \in \bigcup_j M^a_k \cap M^a_l$, where $a$ signifies exclusive consideration of atoms. We optimise both atoms and bonds with a CE loss and denote the combined loss by $L_\sM$.

Next, to establish connectivity on the atom level within the molecular graph, \modelname\ proceeds to identify the \textbf{join nodes \sJ}. The join nodes' types are already determined by the connectivity $A$. \modelname\ accomplishes this by predicting the specific atom positions $p_a$ that need to be combined, collectively forming the join set \sJ. The likelihood of merging nodes $i$ and $j$ in shape representations $M_k$ and $M_l$ is represented by the merge probability $J_{ij}^{(k,l)} = \prob ( p_i \equiv p_j \mid \sM, A, z )$, which constitutes the join matrix $J^{(k,l)}\in~[0,1]^{V_{S_k}\times V_{S_l}}$. It is modelled and optimised using an MLP and CE loss $L_\sJ$.

Finally, predicting the \textbf{leaves $\sL$} involves two key aspects: determining the correct atom type for each leaf and establishing its connection to the molecule's atom representation, denoted as $\sC$ for core molecule. To learn $\prob(\sL_S \mid \sC, z)$, we assess each node position within the shape representation $M_S$. Similar to the approach employed in modelling the shape representation $P(M_i | S, A, z)$ for shape $S_i$, \modelname\ utilises a transformer model to generate atom representations, optimised with CE loss $L_\sL$. This model constructs its memory by encoding the present atom graph and incorporating embeddings tailored to each unique shape representation. The final atom loss is defined by $L_\sG=L_\sM + L_\sJ + L_\sL$.

\subsection{The \modelname\ encoder}
\label{sec:encoder}
\modelname's encoder aims to learn the approximate posterior $\qprob(z\mid \sG)$. At its core, the encoder leverages a graph transformer \textcite{shi_masked_2021} for generating node embeddings of the molecular graph. Since MAGNet generates molecules in a coarse to fine-grained fashion, it is beneficial to encode information about the decomposition of the molecules. To achieve this, we use an additional GNN to capture the coarse connectivity within the shape graph. The embeddings of the molecular and the shape graph are computed by aggregating over the individual atom and shape nodes, respectively. In addition to these aggregations, we exclusively aggregate joins and leaves to represent the other essential components of the graph. The representation of the individual components---molecular graph, shapes, joins, and leaves---are concatenated and mapped to the latent space by an MLP, constituting the graph embedding $z_\sG$. More details about the chosen node features as well as technical specifications of the encoder can be found in \cref{app:encoder_details}.

Taken together, we optimise \modelname\ according to the VAE setting, maximising the ELBO:
\begin{alignat*}{3}
    L &= \mathbb{E}_{z\sim\qprob}\bigl[ \prob\bigl (\sG \mid z\bigr)\bigr] + \beta D_\text{KL}\bigl(\qprob(z\mid \sG) \mid P \bigr) \qquad \text{with} \quad P\sim\mathcal{N}(0,\mathds{1})\\
      &= L_{\sG_\sS} + L_{\sG} + \beta D_\text{KL} \: , 
\end{alignat*}
where the KL-divergence $D_\text{KL}$ serves to regularise the posterior $\qprob(z\mid \sG)$ towards similarity with the Normal prior $P$ in the latent space, weighted by $\beta$. However, we have observed that this regularisation alone is inadequate for achieving a smoothly structured latent space. To address this, we apply a Normalizing Flow to the latent space, aligning it more effectively with the prior \parencite{tong2023improving}.

%% file: chapters/3_relatedwork.tex
\section{Related Work}

\paragraph{Molecule generation}
Existing generative models can be broadly divided into two categories: (1)~string-based models, relying on string representations like SMILES or SELFIES~\parencite{gomez-bombarelli_automatic_2018, segler2018generating, flam-shepherd_language_2022, fang_molecular_2023, adilov_neural_2021, grisoni_chemical_2023}, which do not leverage structural information, and (2) graph-based models, which are inherently centered around molecular graphs.
 Graph-based approaches involve models that represent molecular graphs (1) primarily at the atom level and (2) predominantly through fragments. Regarding the generation process, they can be further divided into sequential methods \parencite{khemchandani_deepgraphmolgen_2020, shi_graphaf_2020, popova_molecularrnn_2019, mercado_graph_2021, luo_graphdf_2021, liu_constrained_2018, li_learning_2018, assouel_defactor_2018, you_graph_2019, yang_hit_2021, lim_scaffold_2020, kajino_molecular_2019, jin_hierarchical_2020, bengio_flow_2021, ahn_spanning_2021, shirzad_tdgen_2022}, building molecules per fragment while conditioning on a partial molecule, and all-at-once (AAO) approaches \parencite{simonovsky_graphvae_2018, ma_constrained_2018, liu_graphebm_2021, decao_molgan_2018, zang_moflow_2020, bresson_two_2019, flam-shepherd_graph_2020, samanta_nevae_2019, kong_molecule_2022b} that create each aspect of the molecular graph in a single step.

\paragraph{Fragmentation and shape representation}
Various techniques are available for constructing fragment vocabularies, with a distinction between chemically-inspired and data-driven approaches. For example, both HierVAE \parencite{jin_hierarchical_2020} and MoLeR \parencite{maziarz_learning_2022} adopt a heuristic strategy known as ``breaking bridge bonds'' to decompose molecules into rings and remainder fragments, emphasising chemically valid substructures. In a similar vein, JT-VAE \parencite{jin_junction_2018} employs fragmentation guided by the construction of junction trees. In contrast, PS-VAE \parencite{kong_molecule_2022b} and MiCaM \parencite{geng2022novo} take a data-driven bottom-up approach, creating fragments by merging smaller components, starting from single atoms. MiCaM even integrates attachment points, resulting in a larger, ``connection-aware'' vocabulary. 

\modelname\ is a graph-based model that employs a unique approach by generating each hierarchy level in a single step, similar to existing AAO approaches. It positions itself between the traditional categories of single-atom and fragment-based models by utilising shapes as building blocks and subsequently generating appropriate atom and bond attributes. Our new fragmentation approach is designed to achieve concise shape representations, expanding its capacity to encode diverse structures within the same vocabulary size.


%% file: chapters/4_experiments.tex
\section{Experiments}

We evaluate \modelname's performance across several dimensions of the generation process. First, we investigate the reconstruction and sampling of shapes $\sS$, as the fundamental component of \modelname's factorisation, and assess how faithfully our model represents the diverse structural characteristics found in molecules. We continue to evaluate the generative performance using established benchmarks. Next, we analyse \modelname's ability to determine suitable atom and bond allocations $\sM$, highlighting its distinctive approach compared to baseline models. Moreover, we demonstrate how \modelname's factorisation and shape vocabulary facilitate its ability to generalise effectively across different datasets in a zero-shot manner. Finally, we explore the possibilities enabled by our factorisation, such as conditioning on various levels of the generation process.
\begin{figure}[t]
\centering
\includegraphics[width=\textwidth]{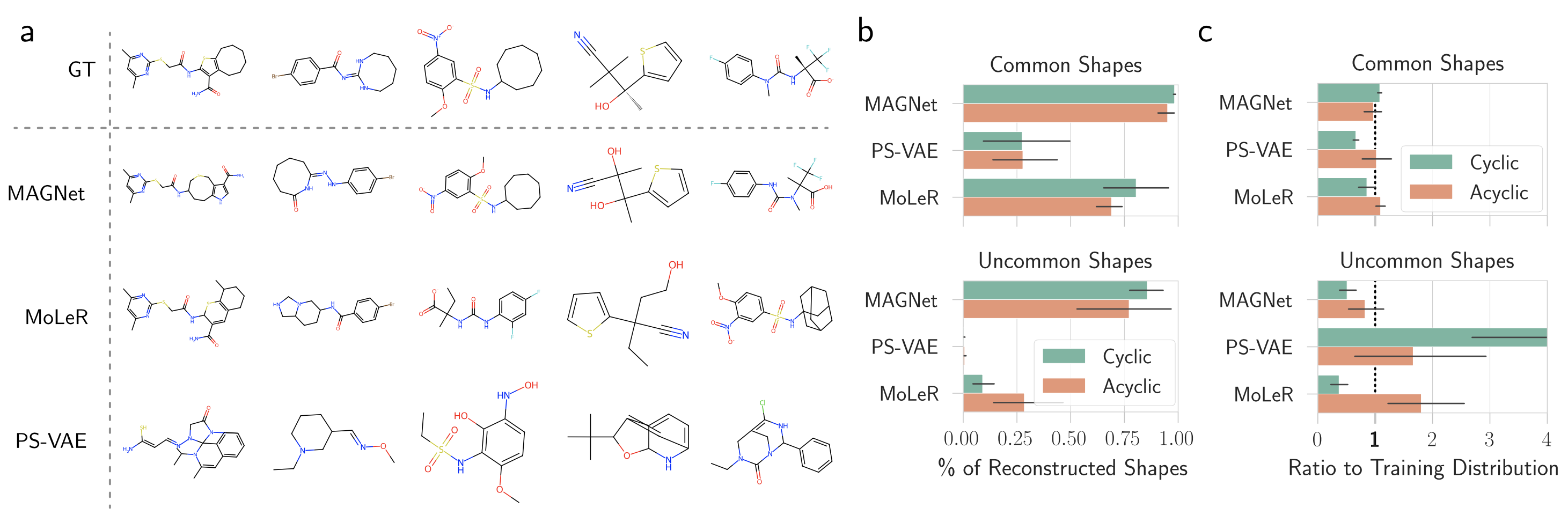}
\caption{(\textbf{a})\,Reconstruction of molecules that include large cycles or complex junctions. Relying on individual atoms to build these structures is not sufficient. Only \modelname\ is able to reliably decode its latent code $z_\sG$. (\textbf{b})\,Percentage of reconstructed shapes. \modelname\ substantially improves in reconstructing both common and---more importantly---uncommon shapes. (\textbf{c})\,Comparison of sampled shapes to shape occurrences in the training distribution. A ratio of $1$ is optimal.\,}
\label{fig:shape_figure}
\end{figure}
\subsection{Using shapes to represent structural variety of molecules}\label{sec:exp_shapes}
\paragraph{Reconstructing large cycles} Our first experiment provides qualitative insights into how accurately shapes are decoded, $\prob(\sS\,|\, z_\sG)$. For this experiment, we employ two baselines: PS-VAE \parencite{kong_molecule_2022b} and MoLeR \parencite{maziarz_learning_2022}, the models with the best performance within their respective category on the GuacaMol Benchmark (see Sec.~\ref{sec:exp_distribution_learning}).
We assess the decoder's performance in reconstructing molecules from the test set, which includes uncommon shapes like large rings or complex junctions. Our observations reveal that the baseline models have difficulty in constructing complex shapes, as illustrated in Figure \ref{fig:shape_figure}a. This limitation is likely attributed to the absence of such shapes in their top-$k$ vocabularies. Consequently, these models face the challenge of constructing shapes such as large rings from individual atoms. In contrast, our proposed model, \modelname, operates with a moderately-sized shape vocabulary that includes complex shapes, enabling it to generate molecules that closely adhere to the latent code and the corresponding ground truth molecules. We quantify this result through the displacement of latent codes in \cref{app:displacement}.
\paragraph{\modelname\ reliably decodes shapes} 
Building on our analysis of large cycles and uncommon junctions, we extend our investigation to assess how effectively different models can reconstruct the shape set $\sS$ in a general context. Given our focus on $\prob(\sS\,|\, z)$, we can disregard the shape connectivity $A$ and representations $\sM$. As illustrated in \cref{fig:shape_figure}b, our findings demonstrate that \modelname\ consistently outperforms both MoLeR and PS-VAE. Notably, our results provide additional support for the hypothesis that the other methods do not effectively learn the concept of a shape, relying primarily on the information encoded in their vocabulary. As a result, they may struggle to decode a common shape $S$ when the signal originally comes from an uncommon representation $M$.
\paragraph{\modelname\ matches the distribution of shapes more accurately} 
%
To further check that uncommon shapes are also sampled, we analyse the shape set of generated molecules. If the other models are able to represent scaffolds that are not included in their vocabulary, they should be able to reflect the reference distribution of shapes. \cref{fig:shape_figure}a shows that this is \emph{not} the case in practice. For this evaluation, we decompose sampled molecules into their shapes. We then measure the models' over- and undersampling behaviour based on the ratio $r_{S_i}$:
\begin{equation*}
    r_{S_i} = \frac{c_{s}(S_i)}{\sum_k c_{s}(S_k)} \times \frac{\sum_k c_{t}(S_k)}{c_{t}(S_i)}\:,
\end{equation*} 
where $c_t$ and $c_s$ refer to the count function applied to the training set and sampled sets, respectively. On common shapes, i.e.\ those that occur in more than 10\% of the molecules, all evaluated models are able to match the ratio of the ZINC distribution. For uncommon shapes, however, both MoLeR and PS-VAE fail: while PS-VAE heavily oversamples both ring-like structures and chain-like structures, MoLeR oversamples chain-like structures and undersamples ring-like structures. \modelname\ matches the reference distribution best across categories and we conclude that the proposed abstraction to shapes is also beneficial for generation.
\subsection{Generative performance evaluated on common benchmarks}
\label{sec:exp_distribution_learning}
\begin{table}[t]
\RawFloats
\centering    
\caption{GuacaMol and MOSES Benchmark. We report mean and standard deviation using 5 random seeds and highlight \underline{the best overall graph-based method} as well as \textbf{the best within each category.}}
\label{tab:distlearning}
\begin{minipage}[t]{0.8\linewidth}
\resizebox{\textwidth}{!}{%
\begin{tabular}{@{} cr| cc |cccc @{}}
&&  \multicolumn{2}{c|}{GuacaMol} & \multicolumn{4}{c}{MOSES} \\
&& FCD $(\uparrow)$ & KL $(\uparrow)$ & IntDiv $(\uparrow)$ & logP $(\downarrow)$ & SA $(\downarrow)$ & QED $(\downarrow)$ \\
\toprule
\multirow{2}*{\rotatebox{90}{SM.}}  
&CharVAE     &        0.17 \scriptsize{$\pm$       0.08} &        0.78 \scriptsize{$\pm$       0.04} &        \textbf{0.88 \scriptsize{$\pm$       0.01}} &                0.87 \scriptsize{$\pm$       0.14} &        0.48 \scriptsize{$\pm$       0.13} &        0.06 \scriptsize{$\pm$       0.03} \\
&SM.-LSTM    &        \textbf{0.93 \scriptsize{$\pm$       0.00}} &        \textbf{1.00 \scriptsize{$\pm$       0.00}} &        0.87 \scriptsize{$\pm$       0.00} &               \textbf{ 0.12 \scriptsize{$\pm$       0.01}} &        \textbf{0.04 \scriptsize{$\pm$       0.02}} &        \textbf{0.00 \scriptsize{$\pm$       0.00}} \\
\midrule
\multirow{5}*{\rotatebox{90}{Sequential}} 
&GraphAF     &        0.05 \scriptsize{$\pm$       0.00} &        0.67 \scriptsize{$\pm$       0.01} &        \textbf{0.93 \scriptsize{$\pm$       0.00}} &                0.41 \scriptsize{$\pm$       0.02} &        0.88 \scriptsize{$\pm$       0.10} &        0.22 \scriptsize{$\pm$       0.01} \\
&HierVAE     &        0.53 \scriptsize{$\pm$       0.14} &        0.92 \scriptsize{$\pm$       0.01} &        0.87 \scriptsize{$\pm$       0.01} &                0.36 \scriptsize{$\pm$       0.17} &        0.20 \scriptsize{$\pm$       0.14} &        0.03 \scriptsize{$\pm$       0.00} \\
&MiCaM       &        0.63 \scriptsize{$\pm$       0.02} &        0.94 \scriptsize{$\pm$       0.00} &        0.87 \scriptsize{$\pm$       0.00} &                0.20 \scriptsize{$\pm$       0.05} &        0.51 \scriptsize{$\pm$       0.03} &        0.08 \scriptsize{$\pm$       0.00} \\
&JTVAE       &        0.75 \scriptsize{$\pm$       0.00} &        0.94 \scriptsize{$\pm$       0.00} &        0.86 \scriptsize{$\pm$       0.00} &                0.28 \scriptsize{$\pm$       0.03} &        0.34 \scriptsize{$\pm$       0.01} &        \underline{\textbf{0.01 \scriptsize{$\pm$       0.00}}} \\
&MoLeR       &       \underline{\textbf{ 0.80 \scriptsize{$\pm$       0.01}}} &        \underline{\textbf{0.98 \scriptsize{$\pm$       0.00}}} &        0.87 \scriptsize{$\pm$       0.00} &               \underline{ \textbf{0.13 \scriptsize{$\pm$       0.02}}} &        \underline{\textbf{0.06 \scriptsize{$\pm$       0.01}}} &        \underline{\textbf{0.01 \scriptsize{$\pm$       0.01}}} \\
\midrule
\multirow{2}*{\rotatebox{90}{AAO}}
&PSVAE       &        0.28 \scriptsize{$\pm$       0.01} &        0.83 \scriptsize{$\pm$       0.00} &        \underline{\textbf{0.89 \scriptsize{$\pm$       0.00}}} &                0.34 \scriptsize{$\pm$       0.02} &        1.18 \scriptsize{$\pm$       0.05} &        0.05 \scriptsize{$\pm$       0.00} \\
&MAGNet      &        \textbf{0.76 \scriptsize{$\pm$       0.00}} &       \textbf{ 0.95 \scriptsize{$\pm$       0.00}} &        0.88 \scriptsize{$\pm$       0.00} &                \textbf{0.22 \scriptsize{$\pm$       0.01}}&        \textbf{0.12 \scriptsize{$\pm$       0.01}} &        \underline{\textbf{0.01 \scriptsize{$\pm$       0.00}}} \\
\bottomrule
\end{tabular}
}
\end{minipage} 
\end{table}
Employing two standard benchmarks for de-novo molecule generation, we establish \modelname's competitive generative performance. The GuacaMol benchmark provides a framework for assessing the ability of a generative model to sample in accordance with the distribution of a molecular dataset \parencite{brown_guacamol_2019}. Next to evaluating the uniqueness and novelty of sampled molecules, the benchmark also computes distributional distances to the reference, i.e.\ the KL-divergence and Fréchet distance (FCD). We use the MOSES benchmark \parencite{polykovskiy_molecular_2020} to report measures for the internal diversity (IntDiv) of generated molecules as well as chemical properties such as synthetic accessability (SA), the octanol-water partition coefficient (logP), and the viability for drugs (QED).

Baselines for these benchmarks additionally include JTVAE \parencite{jin_junction_2018}, HierVAE \parencite{jin_hierarchical_2020}, and MiCaM \parencite{geng2022novo} as sequential, fragment-based methods. For those model with a variable vocabulary, we set the size to 350. We also include GraphAF \parencite{shi_graphaf_2020} as a purely atom-based model. While the focus of this work lies on graph-based molecule generation, we furthermore add the SMILES-based baseline of the GuacaMol benchmark SMILES-LSTM (SM-LSTM) \parencite{segler2018generating} as well as the SMILES-based VAE CharVAE \parencite{gomez2018automatic}. Importantly, SM-LSTM does not have a latent space and can thus not perform targeted decoding. For all baselines, we use the hyperparameters specified in their respective works. We specify \modelname's hyperparameter configuration in \cref{app:hps_training}.
\paragraph{\modelname\ is the best AAO model on standard benchmarks} 
The benchmark is conducted on $10^4$ latent codes sampled from the prior distribution, $z\sim P$, and decoded into valid molecules. Our results for both benchmarks on the ZINC dataset are depicted in \cref{tab:distlearning}. We do not report Novelty and Uniqueness, as almost all evaluated models achieve 100\% on these metrics. Solely GraphAF and HierVAE achieve 91\% and 96\% Uniqueness and Novelty, respectively. For baselines like SM-LSTM and CharVAE, which are not able to achieve 100\%  Validity, we sample until we obtain $10^4$ valid molecules. While MoLeR sets the state of the art on both FCD and KL, \modelname\ overall performs competitively, outperforming all other graph-based baselines. This supports the proposed factorisation in \cref{sec:pG_factorisation} while also challenging the common perception that methods for molecule generation must rely on motif vocabularies to obtain good generative performance.

Moreover, despite the FCD being one of the most important metrics for molecular distribution learning, we find that it fails to provide insights about the structural diversity of the molecules that are generated. Upon further investigation, we evaluate the benchmark on a subset of $10^4$ molecules from the training data. The subset was filtered to include only the 10 most common shapes found in the dataset, resulting in an FCD score of 0.89. This observation offers an explanation for why models like MoLeR can achieve state-of-the-art FCD scores, despite not accurately capturing the distribution of uncommon shapes, as demonstrated in \cref{sec:exp_shapes}. This underscores that our evaluation of the structural diversity of molecule is orthogonal to these benchmarks, providing valuable insights into the tails of the molecular distribution.
\begin{figure}[tbp]
    \includegraphics[width=\textwidth]{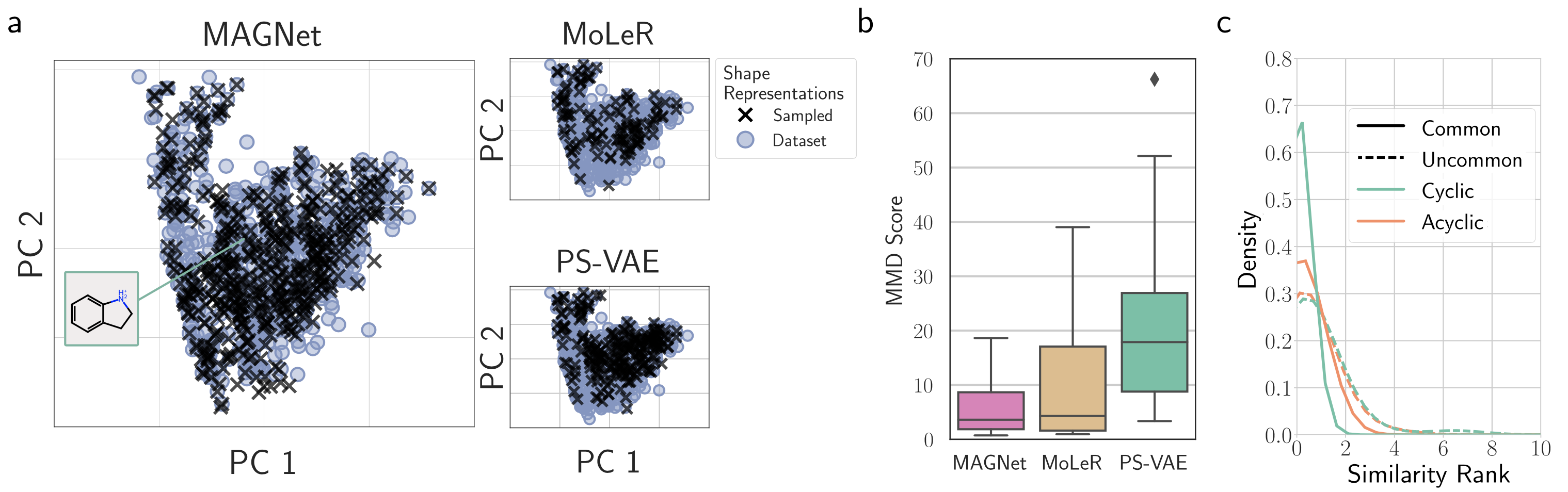}
    \caption{(\textbf{a})\,Example of generated fragments by \modelname\ and baseline methods. (\textbf{b})\,MMD computation to quantify similarity between generated and ground truth shape representations.  (\textbf{c})\,Rank comparison between predicted fragments and their original counterparts.} 
    \label{fig:motif_figure}
\end{figure}
\subsection{Generation of shape representations $\sM$}
Having established \modelname's ability to utilise its shape vocabulary to reliably decode a molecule's structure and sample diversely, we further evaluate \modelname's atom and bond allocation to shapes. 
\label{sec:exp_representation_abiliy}
\paragraph{\modelname's shape representations are superior to fixed fragments}
The larger a given shape, the more the combinatorial aspect starts to dominate: with a size-limited vocabulary, it is challenging to reflect the diversity of a shape's realisations during decoding. This is shown in \cref{fig:motif_figure}a, which provides a qualitative view on shape representations. We extract shape representations of a given shape from the molecules sampled in \cref{tab:distlearning} and plot the two principal components of their fingerprints. Only for this shape, there are 791 representations in ZINC. Both PS-VAE as well as MoLeR are not able to cover the distribution fully, even though the shape appears commonly in the dataset. PS-VAE's and MoLeR's factorisation and vocabulary show lacking ability to cover all shape representations sufficiently. \modelname, by contrast, covers all parts of the distribution, even outliers. \cref{fig:shape_figure}b shows the MMD quantification of the results in \cref{fig:shape_figure}a, confirming that \modelname\ is able to best cover the entire distribution of shape representations. Being able to \emph{reliably} decode a large variety of molecular scaffolds is especially important for downstream tasks such a molecule optimisation. 
\paragraph{Allocation of atom and bonds to shapes} 
Extending the sampling analysis in \cref{fig:shape_figure}b, we quantify the process of turning an abstract shape into a chemically valid substructure in \cref{fig:motif_figure}c. For each shape in the ZINC dataset, we compute the similarities between the set of all predicted and ground truth allocations. Given a ground truth assignment and a successful shape decoding, we measure how the decoded allocation ranks compared to known allocations. In the majority of cases, \modelname\ achieves rank $0$ or $1$ in the shape allocation, with uncommon rings being the most challenging to decode.
\subsection{Application across datasets and conditional sampling}
\label{sec:exp_flexibility}
\begin{figure}[tbp]
    \includegraphics[width=0.9\textwidth]{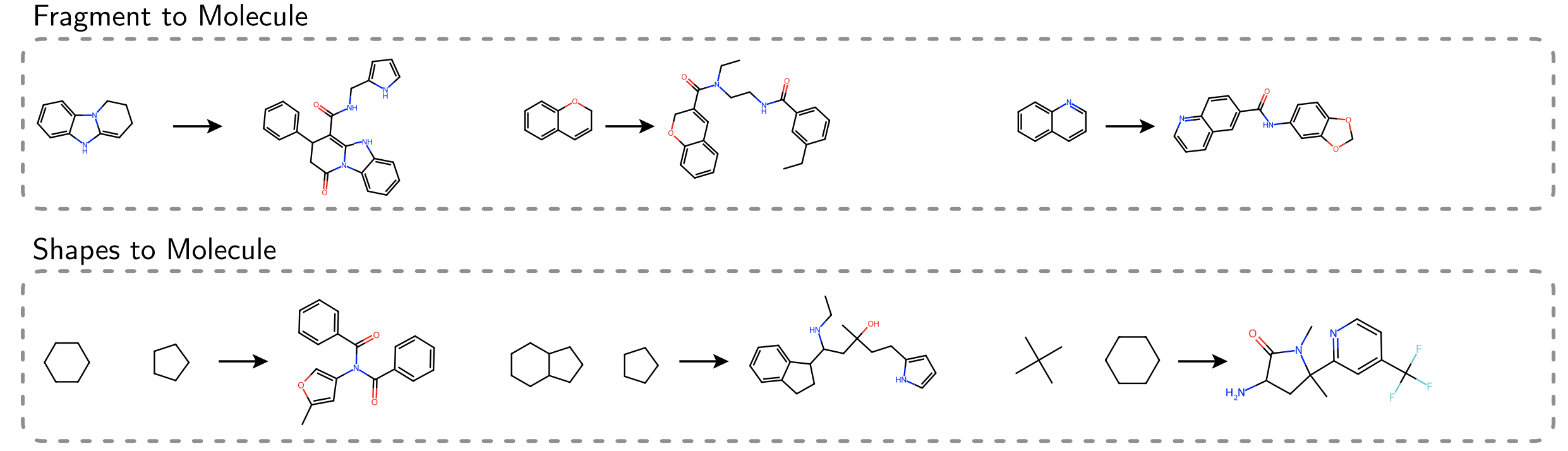}
    \caption{Examples of conditional molecule generation with \modelname. The generation is conditioned on (\textit{top}) a complete fragment, including atoms and edges, and (\textit{bottom}) two distinct shapes.} 
    \label{fig:linker}
\end{figure}

Having analysed the generative performance of the \modelname\ model and the benefit of the proposed shape fragmention, we continue to investigate how well the shape abstraction derived from the ZINC data translates to other datasets. After this, we showcase how one can use \modelname's whole context generation for the generation of linkers and scaffold constrained generation. 
\paragraph{Shape abstractions translate well across datasets}
To examine the flexibility of our shape fragmentation, we evaluate its transferability to unseen datasets with distinct molecular distributions through zero-shot generalization in \cref{app:transferability_vocab}. We use the datasets QM9 \parencite{wu_moleculenet_2018}, GuacaMol \parencite{brown_guacamol_2019}, CheMBL \parencite{mendez2019chembl} and L1000 \parencite{subramanian2017next}.  Note that we do not finetune any model on the unseen datasets and only use the vocabulary extracted from ZINC. \modelname\ is able to achieve the highest similarity scores across all datasets, underscoring the flexibility of the fragmentation and \modelname's expressive power across the space of drug-like molecules.
\paragraph{\modelname\ efficiently generates molecules conditioned on shapes and scaffolds}
In the context of potential downstream applications, we investigate novel scaffold conditioning methods made possible by \modelname's factorisation. Besides the latent space interpolation in \cref{app:interpolation}, Figure \ref{fig:linker} illustrates that \modelname\ is capable to condition not only on a single scaffold but also on multiple scaffolds, even when they are not directly connected within the resulting molecule. This poses a significant challenge for models like MoLeR, which rely on extending connected subgraphs for scaffold conditioning. Moreover, \modelname\ enables conditioning solely based on shapes, enabling the free generation of atoms and bonds---a form of conditioning that was previously not possible. 

%% file: chapters/5_conclusion.tex
\section{Conclusion}

We present \modelname, a generative model for molecules that relies on a novel factorisation to  disentangle structure from features, thus leading to a general abstraction in the space of molecules.
\modelname\ exhibits stronger performance at representing structures than existing models while also showing favourable results in generative tasks.
%
%
While we argue that a global context like the one adopted in \modelname\ is important for shape representations, modifications thereof can also be promising for sequential models.
Finally, our proposed abstraction to shapes lends itself to \emph{general} applications in graph generative models beyond the molecular world.\\[0.4em]
\ifarXiv
\textbf{Acknowledgements\ }
JS and LH are thankful for valuable feedback from David Lüdke, John Rachwan, Tobias Schmidt, Simon Geisler, the DAML group, and Theis Lab. LH is supported by the Helmholtz Association under the joint research school ``Munich School for Data Science - MUDS''.
\fi

%% file: chapters/6_appendix.tex
\newpage
\section{Interpolation}
\label{app:interpolation}

\begin{figure}[t]
    \centering
    \includegraphics[width=\textwidth]{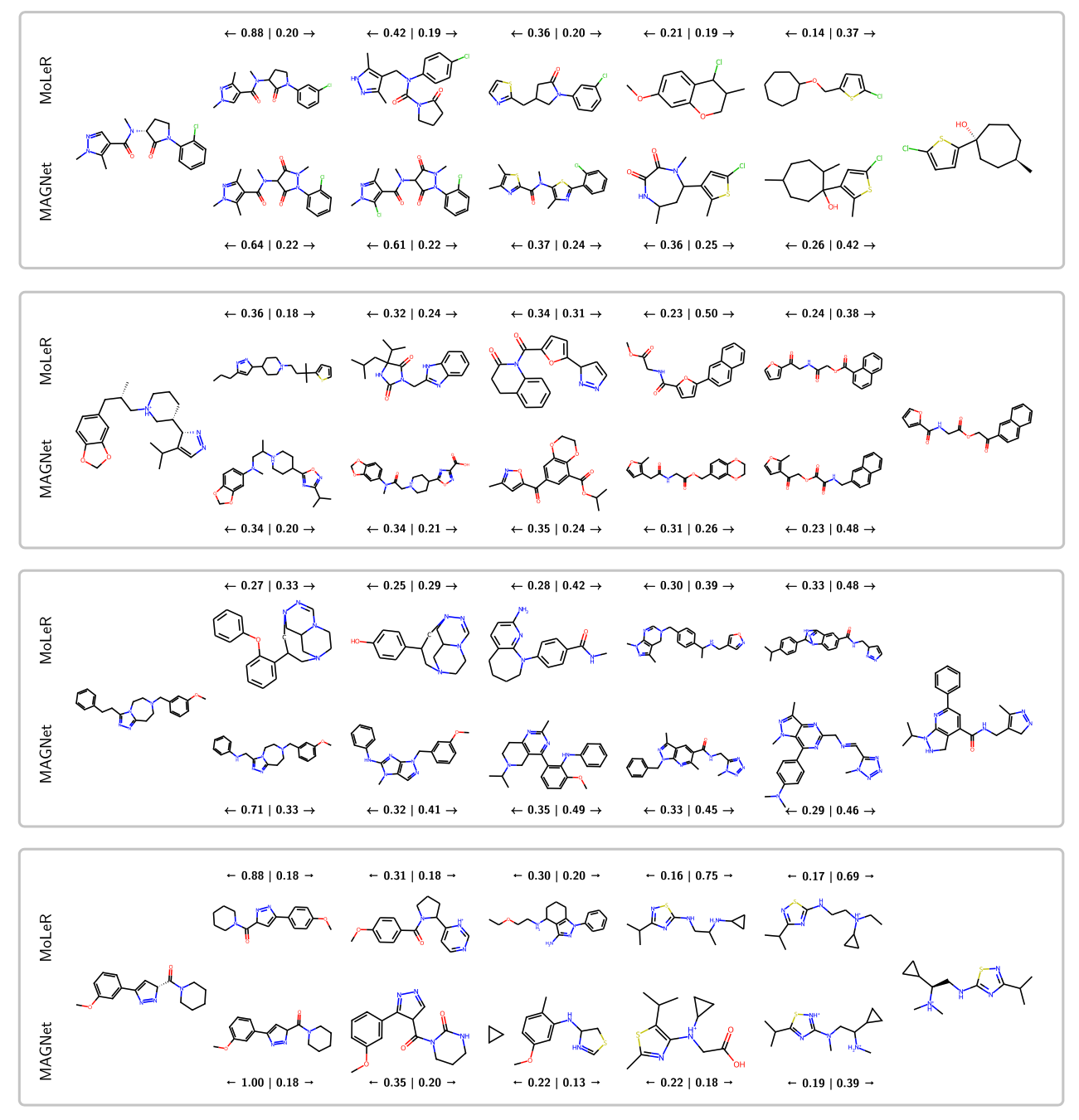}
    \caption{We provide four interpolation examples for MAGNet and MoLeR. The input molecules (left and right) are shared between the two models. We report the Tanimoto similarity as a rough estimate for the interpolation's goodness.}
    \label{fig:app_more_interpolation}
\end{figure}

Extending on \cref{fig:linker}, we additionally provide examples for latent space interpolation in \cref{fig:app_more_interpolation}. During interpolation, \modelname\ stays faithful to the shapes present in the input molecules. The last row shows a failure case of \modelname: it identifies a shape multiset that can not be fully connected to a molecule.

\section{Displacement of Latent Codes}
\label{app:displacement}
To quantify the discrepancy between input and reconstructed molecule visible in \cref{fig:shape_figure}a, we measure the displacement of latent codes. That is, we obtain the latent representation for the input molecule, decode this latent representation into the output molecule and then obtain the latent representation for the output molecule.  This verifies what can be observed qualitatively in \cref{fig:shape_figure}a--the evaluated baselines can not reliably decode complex shapes.

\begin{wrapfigure}[15]{L}{0.5\textwidth}
    \centering
    \includegraphics[width=\textwidth]{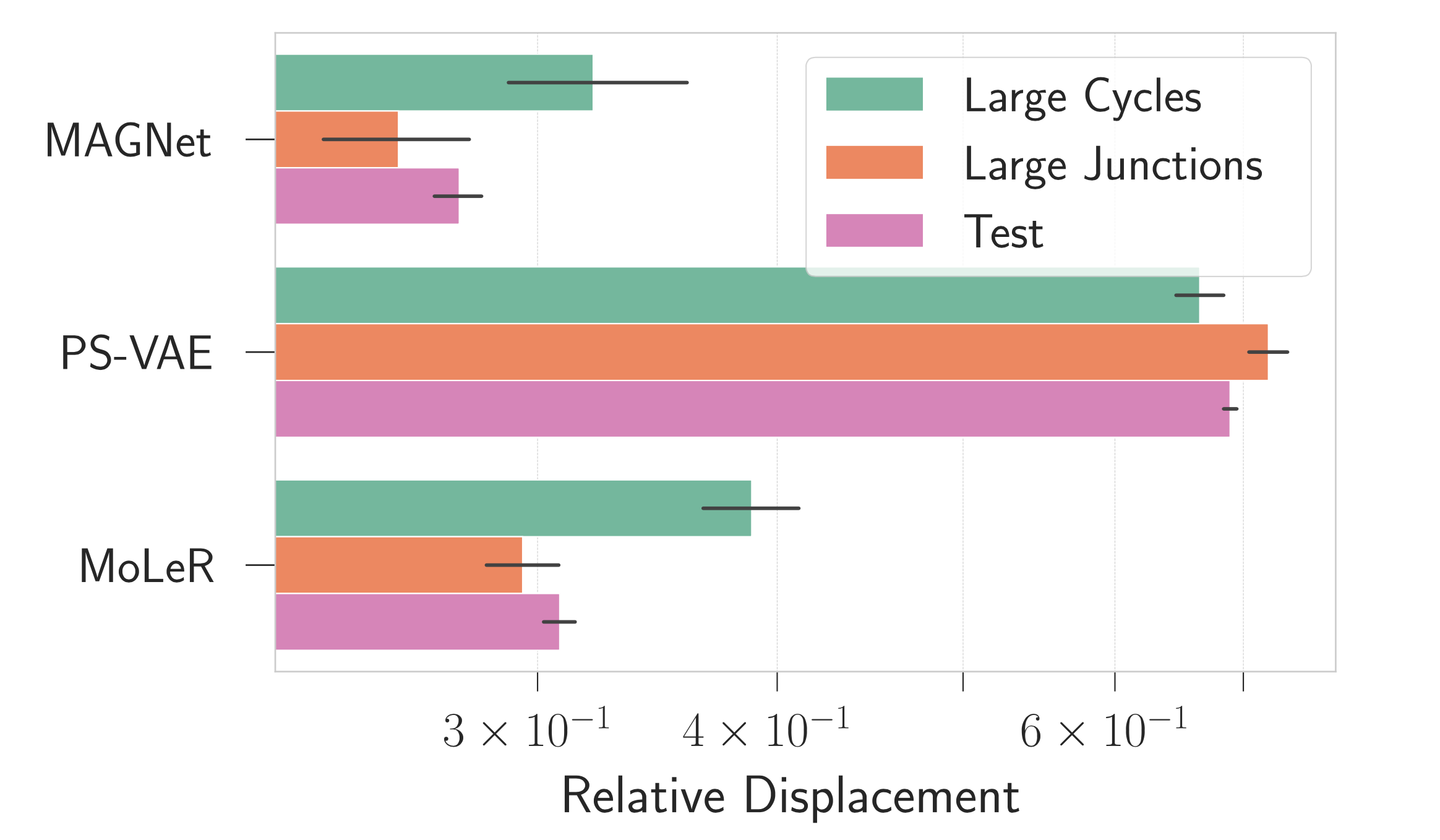}
    \caption{Displacement between latent representation of the input vs. the decoded output.}
    \label{fig:displacement}
\end{wrapfigure}

\section{Transferability of Shapes}
\label{app:transferability_vocab}
\begin{wrapfigure}[15]{r}{0.5\textwidth}
    \centering
    \includegraphics[width=\textwidth]{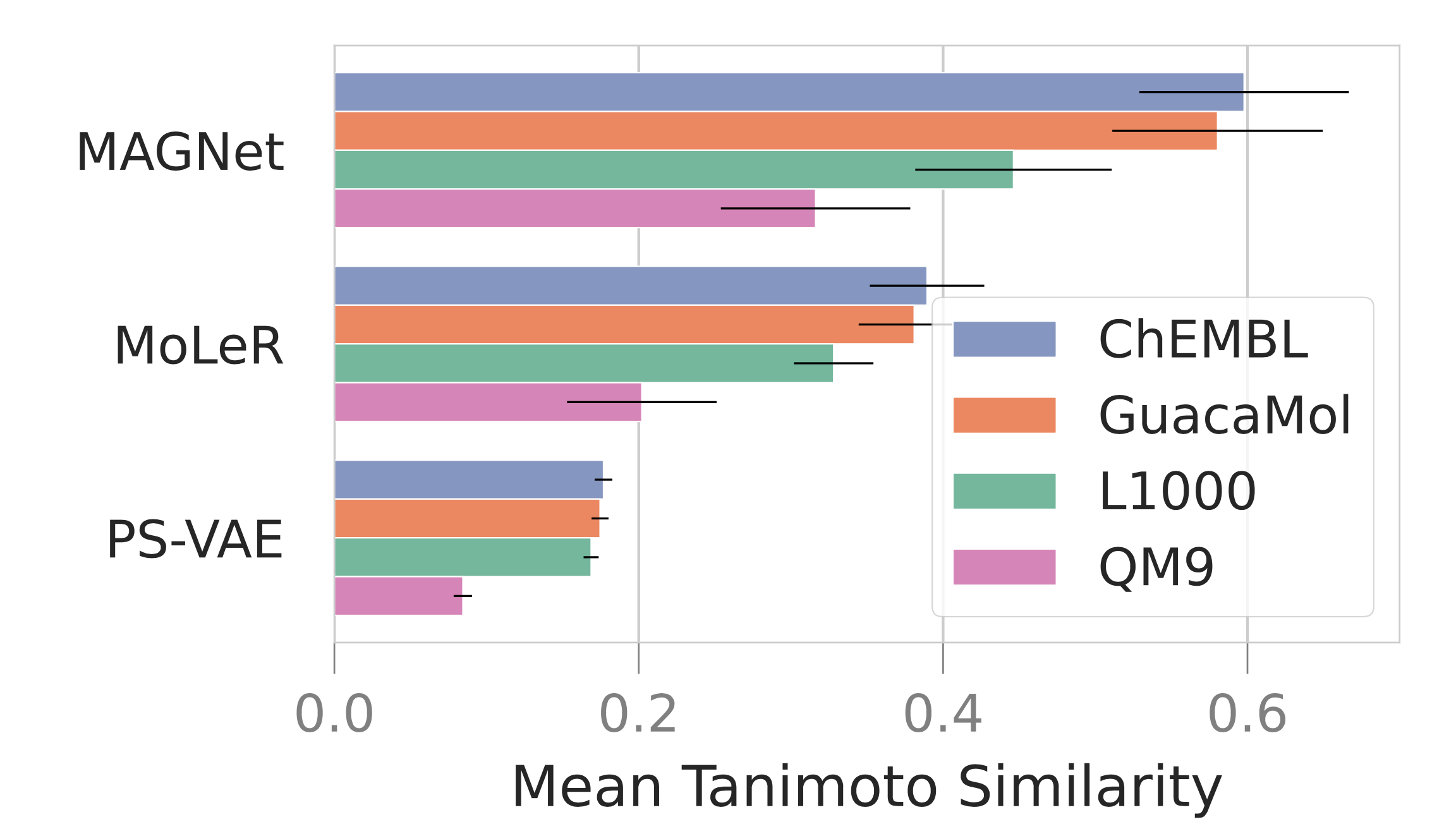}
    \caption{Tanimoto similarity for zero-shot reconstruction on unseen datasets.}
    \label{fig:tanimoto}
\end{wrapfigure}

We calculate the Tanimoto similarity in the reconstruction setting for a variety of datasets, \cref{fig:tanimoto}. For all evaluated datasets, \modelname\ achieves the best similarity scores between molecules, highlighting the transferability of shapes across various distributions.

We compute the tanimoto scores only for those molecules that can be represented via the shapes that were extracted from the ZINC dataset. For the QM9 dataset, \modelname\ can represent roughly 75\% of the molecules in the dataset. This is due to unseen shapes which make up around 11\% out of the total number of 289,966 shapes. For GuacaMol, \modelname\ can represent around 97\% of the molecules in the dataset. Out of the 9,562,028 shapes in GuacaMol, only 0.5\% are missing from the shape vocabulary extracted from the ZINC dataset. We consider a more flexible decomposition of shapes that translates even better across datasets important future work.

\ifarXiv
    \newpage
\else
\fi

\section{Details Encoder}
\label{app:encoder_details}
We build the node features that are processed in \modelname's encoder from different attributes, see \cref{tab:magnet_hyperparams}. We include the atom type (`atom\_id\_dim'), its charge (`atom\_charge\_dim'), as well as its multiplicity value (`atom\_multiplicity\_dim'). We proceed accordingly for the shape level and include the shape id (`shape\_id\_dim'), its multiplicity (`shape\_multiplicity\_dim'), as well chemical features (`motif\_feat\_dim') computed through RDKit \parencite{landrum_rdkit_2010}. Since the latter are not learned during training, the features are mapped to the specified dimensionality by a linear map.

After processing the resulting node features through the graph transformer \parencite{shi_masked_2021} with `num\_layers\_enc'-many layers, they are aggregated in different ways and mapped to specified dimensions as defined by `enc\_<>\_dim' for the atoms, shapes, joins, and leaves, respectively. On top, the shape embeddings are additionally processed with the same transformer architecture (`num\_layers\_shape\_enc') to inform the embedding about the shape-level connectivity. We then concatenate the resulting graph-level embeddings and further combine them with global molecule features, again computed via RDKit and then mapped to the required dimension (`enc\_global\_dim'), before mapping them to the latent space via the latent module which has `num\_layers\_latent'-many layers.

\section{\modelname: Hyperparameters and Training}
\label{app:hps_training}
%
Training \modelname\ for one epoch takes around $30$\,minutes on a single `NVIDIA GeForce GTX 1080 Ti'. We trained \modelname\ for $30$\,epochs and fitted the latent normalizing flow post-hoc for $5000$ epochs in total and conducted a random hyperparameter sweep including the learning rate, beta annealing scheme, and the number of layers for the encoder and latent module. The \modelname\ model reported in the main text has $12.6$\,M parameters and its configuration is depicted in \cref{tab:magnet_hyperparams}. In its current version, \modelname\ processes roughly $70$ molecules per second during training and samples about $8$ molecules per second during inference.
\begin{table}[H]
    \centering
    \begin{tabular}[t]{crll}
    \toprule
    \multicolumn{3}{c}{Parameter}  &                Value \\
    \midrule
    \multicolumn{2}{c}{\multirow{2}{*}{\textbf{Train}}}
    & batch\_size                                             &                   64 \\
    \multicolumn{2}{c}{}& flow\_batch\_size                                  &             1024 \\
    \multicolumn{2}{c}{}& lr                                  &             $3.07\times10^{-4}$ \\
    \multicolumn{2}{c}{}& lr\_sch\_decay                      &                 0.9801 \\
    \multicolumn{2}{c}{}& flow\_lr                                  &             $1\times10^{-3}$ \\
    \multicolumn{2}{c}{}& flow\_lr\_sch\_decay                      &                 0.99 \\
    \multicolumn{2}{c}{}& flow\_patience                      &                 13 \\
    \multicolumn{2}{c}{}& gradclip                            &                    3 \\
    \cmidrule{1-4}
    \multicolumn{2}{c}{\multirow{2}{*}{\textbf{Model}}}
    &  latent\_dim               &                  100 \\
    &\multirow{15}{*}{\rotatebox[origin=c]{90}{dim\_config}}& enc\_atom\_dim             &                   25 \\
    \multicolumn{2}{c}{}& enc\_shapes\_dim           &                   25 \\
    \multicolumn{2}{c}{}& enc\_joins\_dim            &                   25 \\
    \multicolumn{2}{c}{}& enc\_leaves\_dim           &                   25 \\
    \multicolumn{2}{c}{}& enc\_global\_dim           &                   25 \\
    \multicolumn{2}{c}{}& atom\_id\_dim              &                   25 \\
    \multicolumn{2}{c}{}& atom\_charge\_dim          &                   10 \\
    \multicolumn{2}{c}{}& atom\_multiplicity\_dim    &                   10 \\
    \multicolumn{2}{c}{}& shape\_id\_dim             &                   35 \\
    \multicolumn{2}{c}{}& shape\_multiplicity\_dim   &                   10 \\
    \multicolumn{2}{c}{}& motif\_feat\_dim           &                   50 \\
    \multicolumn{2}{c}{}& shape\_hidden         &                  256 \\
    \multicolumn{2}{c}{}& shape\_gnn\_dim            &                  128 \\
    \multicolumn{2}{c}{}& motif\_seq\_pos\_dim &                   15 \\[0.2em]
    \multicolumn{2}{c}{}& leaf\_hidden          &                  256 \\
    \multicolumn{2}{c}{}& latent\_flow\_hidden          &                  512 \\
    \bottomrule
    \end{tabular}
    \begin{tabular}[t]{crll}
    \toprule
    \multicolumn{3}{c}{Parameter}  &                Value \\
    \midrule
    \multicolumn{2}{c}{\multirow{2}{*}{\textbf{Model}}}
    & node\_aggregation                     &                  sum \\
    \multicolumn{2}{c}{}& num\_layers\_latent                   &                    2 \\
    \multicolumn{2}{c}{}& num\_layers\_enc                      &                    2 \\
    \multicolumn{2}{c}{}& num\_layers\_shape\_enc               &                    4 \\
    \multicolumn{2}{c}{}& num\_layers\_hgraph                   &                    3 \\[1.5em]
    \cmidrule{3-4}
    &\multirow{4}{*}{\rotatebox[origin=c]{90}{loss\_weights}}
    & joins                  &                    1 \\[0.2em]
    \multicolumn{2}{c}{}& leaves                 &                    1 \\[0.2em]
    \multicolumn{2}{c}{}& motifs                 &                    1 \\[0.2em]
    \multicolumn{2}{c}{}& hypergraph             &                    1 \\[1.5em]
    \cmidrule{3-4}
    & \multirow{5}{*}{\rotatebox[origin=c]{90}{beta\_annealing}}
    &  max                  &                    0.01 \\[0.2em]
    \multicolumn{2}{c}{}& init                 &                 0 \\[0.2em]
    \multicolumn{2}{c}{}& step                 &               0.0005 \\[0.2em]
    \multicolumn{2}{c}{}& every                &                  2500 \\[0.2em]
    \multicolumn{2}{c}{}& start                &                 2000 \\[7.48em]
    \bottomrule

    \end{tabular}
    \caption{Parameter configuration of the best \modelname\ runs.}
    \label{tab:magnet_hyperparams}
\end{table}